\documentclass[preprint,aps]{revtex4}
\usepackage{graphicx}
\usepackage{epsfig}
\usepackage{epstopdf}
\usepackage{dcolumn}% Align table columns on decimal point

\usepackage{bm}% bold math
\usepackage{amsmath}% AMS math commands
\usepackage{amssymb}
\usepackage{placeins} %to use \FloatBarrier command

\def\be{\begin{equation}}
\def\ee{\end{equation}}
\def\bea{\begin{eqnarray}}
\def\eea{\end{eqnarray}}

\def\ua{\uparrow}
\def\da{\downarrow}

\def\uup{{u^\uparrow}}
\def\udn{{u^\downarrow}}
\def\dup{{d^\uparrow}}
\def\ddn{{d^\downarrow}}

\begin{document}

\title{Re-examining valence quark spin distributions}

%\pacs{12.38.Aw, 12.39.Ba, 14.20.Dh }
%\keywords      {Quark distributions, Quark model, Nucleon}

\author{A. I. Signal}
\email{a.i.signal@massey.ac.nz}{
\affiliation{Institute of Fundamental Sciences PN461 \\ Massey University \\ Palmerston North 4442 \\ New Zealand}

\vskip0.5cm

\begin{abstract}

The observed deep inelastic nucleon spin asymmetries $A_{1}^{p,n}$ and the spin dependent quark distributions break SU(6) spin - flavour symmetry. 
In quark models two mechanisms for breaking SU(6) symmetry are well known: the hyperfine interaction and the pion cloud of the nucleon.
I re-evaluate these mechanisms and show how the breaking of SU(6) in each case affects spin dependent valence quark distributions of the proton.
%In particular I investigate the large $x$ properties of these mechanisms and attempt to test these against known data. 
In particular I investigate the properties of these mechanisms in the kinematic region $0.3 < x < 0.7$ and attempt to test these against known data. 
Both mechanisms are able to explain the quantitative features of the spin asymmetries.

\end{abstract}

%\pacs{12.38.Aw, 12.39.Ba, 14.20.Dh }
%\keywords{Quark distributions, Quark model, Nucleon}

\maketitle

%%%%%%%%%%%%%%%%%%%%%%%%%%%%%%%%%%%%%%%%%%%%
%% MAINMATTER
%%%%%%%%%%%%%%%%%%%%%%%%%%%%%%%%%%%%%%%%%%%%

\section{Introduction}

Does the quark model of nucleons offer any insight into the quark distribution functions observed at large momentum transfer? 
Nearly 30 years have passed since the EMC data on the spin carried by quarks in deep inelastic scattering was published \cite{EMC88}, 
and this question has yet to be answered definitively. 
In a 1999 paper, Isgur wrote {\em it is surprising that we still do not know whether our \ldots picture of the spin structure of the valence quarks is right} \cite{Isgur99}, 
and the situation is little improved now.
In this paper I re-evaluate Isgur's work and review progress on relating quark model predictions to spin dependent quark distribution functions.
I firstly study the role of the hyperfine interaction in breaking SU(6) symmetry and whether this interaction plays an important part in the 
valence quark distributions at large $x$. 
In addition I investigate whether the pion cloud of the nucleon, which is required from chiral symmetry considerations in many models, has any effect 
on these distributions in the large $x$ region. 
I calculate the relative importance of these two mechanisms for SU(6) breaking using the MIT bag model and the cloudy bag model, and show that in this 
kinematic region the pion cloud has the larger effect, and moreover is able to reproduce many of the known features of the experimental data.

\section{SU(6) symmetry predictions}

The SU(6) wavefunction of the proton may be written as 
\bea
|p \ua \rangle = \frac{1}{\sqrt{2}} \left [ \phi_{MS}(uud) \chi_{MS}(\ua \ua \da) + \phi_{MA}(uud) \chi_{MA}(\ua \ua \da) \right ] |\psi_{0} \rangle
\label{eq:pwf}
\eea
where $\phi_{MS}$ and $\phi_{MA}$ are the mixed symmetry representations of SU(3) flavour which are symmetric or anti-symmetric respectively in the first pair of quarks, 
and similarly $\chi_{MS}$ and $\chi_{MA}$ are the mixed symmetry spin wave functions. 
The spatial wavefunction $\psi_{0}$ is assumed to be symmetric and have angular momentum $L = 0$.
Here I have suppressed colour indices, but an overall anti-symmetrization of the wavefunction due to colour is assumed.

There will be a quark valence distribution $v(x)$ associated with $\psi_{0}$ and its light-cone properties, which will be the same for each of the three quarks, 
so the spin dependent valence distributions are 
\bea
\udn(x) & = & \frac{1}{3} v(x) = \frac{1}{5} \uup(x) \\
\dup(x) & = & \frac{1}{3} v(x) = \frac{1}{2} \ddn(x).
\eea
These distributions predict 
\bea
\frac{F_{2}^{n}(x)}{F_{2}^{p}(x)} & = & \frac{4d(x) + u(x)}{4u(x) + d(x)} = \frac{2}{3} \\
A_{1}^{p} & = & \frac{4\Delta u(x) + \Delta d(x)}{4u(x) + d(x)} = \frac{5}{9} \\
A_{1}^{n} & = & \frac{4\Delta d(x) + \Delta u(x)}{4d(x) + u(x)} = 0.
\eea

This ignores relativistic effects in the spatial wavefunction. 
The lower component of the quark spinor converts around 25\% of the quark spin into orbital angular momentum.
If the probability of a quark to have its spin opposite to its total angular momentum is denoted $\frac{1}{2}c_{A}(x)$, then 
\bea
\Delta u(x) & = & \frac{4}{3}(1 - c_{A}(x))v(x)  \nonumber  \\
\Delta d(x) & = & -\frac{1}{3}(1 + c_{A}(x))v(x).
\eea
This gives the relation for the integrated distributions 
\bea
\Delta u - \Delta d = \frac{5}{3} - \langle c_{A}(x) v(x) \rangle
\eea
which will agree with the Bjorken sum rule for reasonable spin flip probabilities, such as in the bag model.
The relativistic correction also changes the SU(6) prediction for the proton spin asymmetry 
\bea
A_{1}^{p} = \frac{5}{9} [1 - c_{A}(x)].
\eea

\section{Hyperfine Interaction}

The hyperfine (spin - spin) interaction arising from one gluon exchange is important in quark models.
Among the well-known effects of the hyperfine interaction are the splitting of the nucleon and delta masses, the negative charge radius of the neutron, 
and violations of SU(6) selection rules \cite{IKK78, Close79}.
For the nucleon, only the contact term of the hyperfine Hamiltonian --- $\vec{S_{i}} \cdot \vec{S_{j}} \delta^{3}(\vec{r_{ij}})$ --- contributes. 
In the rest frame this raises the energy of pairs of quarks with spin 1, and lowers the energy of pairs with spin zero.
The problem here is that the spin components $\chi_{MS}$, $\chi_{MA}$ are not eigenfunctions of the hyperfine interaction.
There are two approaches that can be taken: 1) attempt an exact solution in terms of eigenfunctions of the hyperfine interaction, or 2) use a perturbative approach.
The exact solution follows the arguments of Close \cite{Close73} and Carlitz and Kaur \cite{CK77}, whereas the perturbation approach is used by Isgur \cite{Isgur99}.

\subsection{Hyperfine Basis}

The proton wavefunction (\ref{eq:pwf}) can be written in terms of hyperfine eigenfunctions as 
\bea
|p \ua \rangle & = & \left [ \frac{1}{\sqrt{2}} u \!\ua  \phi_{S}(ud) \chi_{0} + \frac{1}{\sqrt{18}} u \!\ua \phi_{A}(ud) \chi_{10} - \frac{1}{3} u \!\da \phi_{S}(ud) \chi_{11}  \right.  \nonumber \\
&&\;\; \left. \ - \frac{1}{3} d \!\ua \phi_{S}(uu) \chi_{11} + \frac{\sqrt{2}}{3} d \!\da \phi_{S}(uu)\chi_{11} \right ] |\psi_{0} \rangle.
\label{eq:pwhf}
\eea
where $\phi_{S}$,  $\phi_{A}$ are the symmetric and antisymmetric flavour wavefunctions, and $\chi_{0}$, $\chi_{1m}$ are the singlet and triplet (with projection $m$) spin wavefunctions.
Turning on the hyperfine interaction raises the degeneracy of the singlet and triplet spin components, and explicitly breaks SU(2) spin symmetry. 
Note that the wavefunction is still symmetric under SU(3) flavour.
If $v_{0}(x)$ and $v_{1}(x)$ are the valence distributions associated with the $S = 0$ and $S = 1$ spectators, we have the spin dependent valence distributions \cite{CT88}:
\bea
\uup(x) & = & \frac{3}{2} v_{0}(x) + \frac{1}{6} v_{1}(x) \\
\udn(x) & = & \frac{1}{3} v_{1}(x) \\
\dup(x) & = & \frac{1}{3} v_{1}(x) = \frac{1}{2} \ddn(x).
\eea
The peaks of these distributions are located near $ x = 1 - m_{s}/m_{p} $, where $m_{s}$ is the `mass' of the spectator diquark \cite{CT88, SST91}, 
so the hyperfine interaction pushes $v_{1}(x)$ to lower $x$, and $v_{0}(x)$ will dominate at large $x$.
This leads to the predictions for the large $x$ behaviour of the structure functions and asymmetries: 
\bea
\frac{F_{2}^{n}(x)}{F_{2}^{p}(x)} & \rightarrow &  \frac{1}{4} \\
A_{1}^{p} & \rightarrow & 1 \\
A_{1}^{n} & \rightarrow & 1
\eea
where I have assumed that the relativistic spin flip probability $\frac{1}{2}c_{A}(x) \rightarrow 0$ at large $x$.
These predictions are substantially different from the SU(6) predictions, but are built in to most phenomenological fits to experimental parton distributions.

I note that the approach used here is similar in spirit to the method usually used to calculate the hyperfine splittings of the $N - \Delta$ and $\Sigma - \Lambda$ in models of 
hadrons eg.\ the non-relativistic quark model \cite{DeRGG75}, or the MIT bag model \cite{MIT75}. 
In particular the spatial wavefunction of the quarks is the same for all terms in the wavefunction regardless of whether the quark pairs are in singlet or triplet states, 
and the mass splitting is entirely due to the difference in energy between the singlet and triplet spin wavefunctions.
This is similar to the case of hyperfine corrections in atomic physics, where, for instance, the $F = 0$ and $F = 1$ states in the ground state of Hydrogen have the same spatial wavefunction. 
Now we could write the proton wavefunction in the hyperfine basis as 
\bea
|p \ua \rangle & = & \frac{1}{\sqrt{2}} u \!\ua  \phi_{S}(ud) \chi_{0} |\psi_{0} \rangle + \left [ \frac{1}{\sqrt{18}} u \!\ua \phi_{A}(ud) \chi_{10} - \frac{1}{3} u \!\da \phi_{S}(ud) \chi_{11}  \right.  \nonumber \\
&&\;\; \left. \ - \frac{1}{3} d \!\ua \phi_{S}(uu) \chi_{11} + \frac{\sqrt{2}}{3} d \!\da \phi_{S}(uu)\chi_{11} \right ] |\psi_{1} \rangle
\label{eq:2chf}
\eea
where now $\psi_{1}$ describes the spatial wavefunction of the $S = 1$ spectators plus the struck quark.
In the papers of Close \cite{Close73} and Carlitz and Kaur \cite{CK77}, it is assumed that breaking SU(6) by the hyperfine interaction requires $\psi_{0} \neq \psi_{1}$. 
However, in a model context $\psi_{1}$ would describe an excited state of the model Hamiltonian, which is typically $200 - 300$ MeV more massive than the ground state $\psi_{0}$, 
and would not have the same symmetry properties as the ground state (eg.\  having negative parity).
Also, as pointed out in \cite{Isgur99}, the wavefunction in equation (\ref{eq:2chf}) is inconsistent with the Pauli principle when $\psi_{1} \neq \psi_{0}$ unless the wavefunctions have special permutation 
properties, or the full wavefunction is antisymmetrized. 
In one sense the difficulty with trying to take $\psi_{1} \neq \psi_{0}$ is that the {\em ansatz} of an exact solution in terms of the eigenfunctions of the hyperfine interaction assumes that 
there are no other terms in the Hamiltonian (or that the hyperfine states are eigenstates of any extra terms), which can only be true of fairly simple models.
 
I note that an attempt to implement an antisymmetrized wavefunction with hyperfine correction was made in \cite{PTB02}. 
Using the Isgur and Karl model \cite{IK78} in a light-front formalism, they showed that the unpolarized ratio $F_{2}^{n} / F_{2}^{p}$ was larger than the SU(6) value of $\frac{2}{3}$ 
for $x > 0.5$, even though the experimental data are much smaller than $\frac{2}{3}$ in this region.

\subsection{Hyperfine Perturbation}

In a perturbative approach, the physical nucleon is a mixture of the pure SU(6) eigenstates of the nucleon and appropriate excited states. 
At lowest order $L = 2$ and antisymmetric $L = 0$ contributions can be ignored, and the dominant contribution comes from the $N(70, 0^{+}) (1710)$ \cite{IKK78}. 
The nucleon wavefunction then is 
\bea
| \tilde{N} \rangle \simeq | N_{S} \rangle \cos \theta + | N_{M} \rangle \sin \theta 
\eea
where $| N_{S} \rangle$ can be taken as a mixture of 56 symmetric states, dominated by the SU(6) nucleon wavefunction, and $| N_{M} \rangle$ is the SU(6) octet state \cite{Close79, IK79}
\bea
| 8^{2}(70, 0^{+}) \frac{1}{2}^{+} \rangle = \frac{1}{2} \left [ (\phi_{MA} \chi_{MA} - \phi_{MS} \chi_{MS}) | \psi_{0}^{'} \rangle +  (\phi_{MA} \chi_{MS} + \phi_{MS} \chi_{MA}) | \psi_{0}^{''} \rangle \right ].
\label{eq:nstar}
\eea
Here the spatial wavefunctions $\psi_{0}^{'}, \psi_{0}^{''}$ are appropriate excited states of the model Hamiltonian being used, still with $L=0$.

I note that the wavefunctions I am using for excited states are those of the conventional SU(6) quark model, which allows for direct comparison with the work of  \cite{Isgur99}.
There is now the strong possibility that the excited states' wavefunctions have large pentaquark components \cite{JW03} eg.\ the $N(1710)$ may correspond to the state $|[ud][su]\bar{s} \rangle$, 
which would lead to different effects on the valence distributions. 
This will be considered in future work.
 
%Isgur \cite{Isgur99} takes a mixing angle of $\sin \theta  \simeq -0.23$, slightly smaller than the value quoted in \cite{IKK78}.
Taking the mixing angle to be small, at lowest order the perturbed proton wavefunction becomes:
\bea
|\tilde{p} \ua \rangle & = & \frac{\sqrt{2}}{6} [ 2 ( u \!\ua d \!\da u \!\ua + d \!\da u \!\ua u \!\ua + u \!\ua u \!\ua d \!\da) - \nonumber \\ 
& & \; (u \!\da d \!\ua u \!\ua + u \!\ua d \!\ua u \!\da + d \!\ua u\! \da u \!\ua + d \!\ua u \!\ua u \!\da + u \!\ua u \!\da d\! \ua + u \!\da u \!\ua d \!\ua ) ] |\psi_{0} \rangle \nonumber \\
& & +  \frac{\theta}{6}  [ (u \!\ua d \!\da u \!\ua + d \!\da u\! \ua u \!\ua + u \!\ua u \!\da d \!\ua + u \!\da u \!\ua d \!\ua + u \!\ua d \!\ua u \!\da + d \!\ua u \!\ua u \!\da) - \nonumber \\
& & \;\; 2 (d \!\ua u \!\da u \!\ua + u \!\da d \!\ua u \!\ua + u \!\ua u \!\ua d \!\da ) ] |\psi_{0}^{'}  \rangle \nonumber \\
& & + \frac{\sqrt{3}\theta}{6} [ u \!\ua d \!\da u \!\ua - d \!\da u \!\ua u \!\ua - u \!\ua u \!\da d \!\ua + u \!\da u \!\ua d \!\ua - u \!\ua d \!\ua u \!\da + \nonumber \\
& & \;\; d \!\ua u \!\ua u \!\da ] | \psi_{0}^{''}  \rangle . 
\label{eq:hpwf}
\eea
In the $uds$ basis of \cite{Isgur99, IK79} the $\psi_{0}^{''}$ (labelled $\psi^{\rho}$ in \cite{Isgur99} and $\psi_{00}^{\rho}$ in \cite{IK79}) 
piece of the wavefunction makes no contribution to the $d$ distributions.
However, this is not the case in the SU(6) basis used here, where both $\psi_{0}^{'}$ and $\psi_{0}^{''}$ contribute to the $u$ and $d$ distributions. 
At lowest order, the perturbation gives rise to interference terms arising from matrix elements $ \langle \psi_{0}^{'} | \hat{H}_{\rm hf} | \psi_{0} \rangle $ and $ \langle \psi_{0}^{''} | \hat{H}_{\rm hf}  | \psi_{0} \rangle $. 
These distort the spin-averaged SU(6) distribution $v(x)$ oppositely for $u$ and $d$ quarks:
\bea
u_{v}(x) & = & 2v(x) + \theta w(x) \nonumber \\
d_{v}(x) & = & v(x) - \theta w(x)
\eea
where $w(x)$ is the distortion introduced by the hyperfine interaction.
I obtain the same spin dependent distributions as \cite{Isgur99} when the relativistic correction is also applied:
\bea
\uup(x) & = & [1 - \frac{1}{2} c_{A}(x)] u_{v}(x) - \frac{1}{3}[1 -  c_{A}(x)] d_{v}(x) \\
\udn(x) & = & \frac{1}{2} c_{A}(x) u_{v}(x)  + \frac{1}{3}[1 -  c_{A}(x)] d_{v}(x) \\
\dup(x) & = & \frac{1}{3}[1 +  \frac{1}{2} c_{A}(x)] d_{v}(x)  \\
\ddn(x) & = & \frac{2}{3}[1 -  \frac{1}{4} c_{A}(x)] d_{v}(x).
\eea
If the distortion term $\theta w(x)$ is positive at large $x$, then $u_{v}(x)$ dominates, similar to the results of the calculation using the hyperfine basis. 
However, there is no reason for the sign of the distortion to be positive or negative for any value of $x$ --- clearly  the integral of $w(x)$ over $x$ must vanish, 
but there is no further general information on the sign of the distortion. 
In \cite{Isgur99}, it is assumed that the distortion of the valence distributions gives a similar dominance by the $u_{v}(x)$  distribution as in the hyperfine basis, 
and the figures 1(a) and 1(b) of that paper show the spin asymmetries for a reasonable range of parameters for the relativistic correction $\frac{1}{2} c_{A}(x)$ 
and the large $x$ behaviour of $u_{v}(x)$ and $d_{v}(x)$. 

I note that the structure function ratio 
\bea
\frac{F_{2}^{n}(x)}{F_{2}^{p}(x)}  =  \frac{6v(x) + 5 \theta w(x)}{9v(x) + 5 \theta w(x)} 
\eea
can be greater or less than the SU(6) value of $\frac{2}{3}$ depending on the sign of $\theta w(x)$.
Also it is necessary that $|w(x) / v(x)| \sim O(1)$ for the corrections to the SU(6) symmetry predictions to be large.

\subsection{Hyperfine Perturbation in the MIT bag model}

At this stage it is interesting to investigate the mixing between the 56 symmetric nucleon state and the 70 mixed state in the bag model.
This allows a comparison with the calculation in the non-relativistic quark model, and may give insight into which features of the perturbation arising from the hyperfine interaction 
are model independent.

The conventional SU(6)$\times$O(3) $N(70, 0^{+})$ wavefunction in the bag model is \cite{Close79, DeGJ76, CH80, UDW83}
\bea
| N^{*}(70) \rangle = \frac{1}{2} \left [ (\phi_{MA} \chi_{MA} - \phi_{MS} \chi_{MS}) | \psi_{MS} \rangle +  (\phi_{MA} \chi_{MS} + \phi_{MS} \chi_{MA}) | \psi_{MA} \rangle \right ].
\label{eq:nstarbag}
\eea
where the spatial wavefunctions are now mixed symmetry representations of two quarks in the ground ($1s$) state and one quark in the $2s$ excited state. 
These bag model states for bag radius $R$ are 
\bea
\psi_{n}(\mathbf{r}) = N_{n} \left( \begin{array}{c} j_{0}(\omega_{n} r ) \\ i \vec{\sigma} \cdot \hat{\mathbf{r}} j_{1}(\omega_{n} r) \end{array} \right) \chi \Theta(R - r) 
\label{eq:bagwf}
\eea
where $\chi$ is a Pauli spinor, $n$ is the radial quantum number and $\omega_{n}$ is the eigenenergy:
\bea 
\Omega_{n} = \omega_{n}R = \left\{ \begin{array}{ll} 2.04 \ldots & n = 1 \\ 5.40 \ldots & n=2 \end{array} \right. . 
\eea
The normalisation is given by 
\bea
N_{n}^{2} = \frac{1}{4 \pi R^{3}} \frac{\Omega_{n}^{2}}{1 - j_{0}(\Omega_{n})}.
\eea
We can estimate the masses of the baryons by minimising the total energy of the bag state, being the sum of the volume energy, the quark eigenenergies and the zero point energy,
\bea
E(R) = \frac{4}{3} \pi R^{3} B + \frac{1}{R} \sum_{i} \Omega_{i} - \frac{Z_{0}}{R}
\eea
with respect to variations of the bag radius $R$.
I use values determined from previous studies \cite{DeGJ76, DJ80, DeTar81} of $B^{1/4} = 0.15$ GeV and $Z_{0} = 1.0$ GeV fm.
This gives 
\bea
E_{N} = 1.28 \: {\rm  GeV} & R_{N} = 5.33 \: {\rm  GeV}^{-1} \; (1.05 \; {\rm fm})  \nonumber \\
E_{N^{*}} = 1.87 \: {\rm  GeV} & R_{N^{*}} = 6.04 \: {\rm  GeV}^{-1} \; (1.19 \; {\rm fm}) .
\eea
These values include spurious centre-of-mass motion, which needs correction. 
The momentum of the centre-of-mass is approximately \cite{DJ80, Wong81}
\bea
\langle P_{\rm cm}^{2} \rangle \approx \sum_{i} \left( \frac{\Omega_{i}}{R} \right)^{2}
\eea
and the baryon masses are then 
\bea
M = \sqrt{E^{2} - \langle P_{\rm cm}^{2} \rangle}, 
\eea
which results in 
\bea
M_{N} & = & 1.09 \: {\rm  GeV} \\
M_{N^{*}} & = & 1.57 \: {\rm  GeV}.
\label{eq:mass}
\eea

Treating one gluon exchange as a perturbation to the bag Hamiltonian will give rise to a mixing of the degenerate $N^{*}(56)$ and $N^{*}(70)$ states \cite{CH80, UDW83}, 
and mix these states with with the nucleon. 
The mixing with the nucleon arises from the diagram shown in figure 1, where a $1s$ quark is excited to a $2s$ state, and the contribution is denoted $c^{2,1}$ by 
Close and Horgan \cite{CH80}.
This only occurs because the bound quarks are off-shell.
I find $c^{2,1} = 2.36 \times 10^{-4}$, and obtain the off-diagonal matrix elements 
\bea
\langle N^{*} | \hat{H}_{\rm hf} |N \rangle&  = & \frac{16}{3} \frac{\alpha_{s}}{R} \sum_{i>j} (\sigma_{i} \cdot \sigma_{j}) c^{2,1} \nonumber \\
& \approx & - 1 \: {\rm  MeV}.
\eea
Here the strength of the strong interaction in the bag is set by the $N - \Delta$ splitting, which gives \cite{CH80}
\bea
\kappa = \frac{8 \alpha_{s}}{3R}  = 280 \: {\rm  MeV}.
\eea
This results in a mixing angle between the nucleon and the $N^{*}(70)$ of $\theta \approx +0.002$. 
This is in contrast to the mixing angle of $-0.23$ in the non-relativistic quark model.

\begin{figure}[bt]
\centering
  \includegraphics[width=8.6cm]{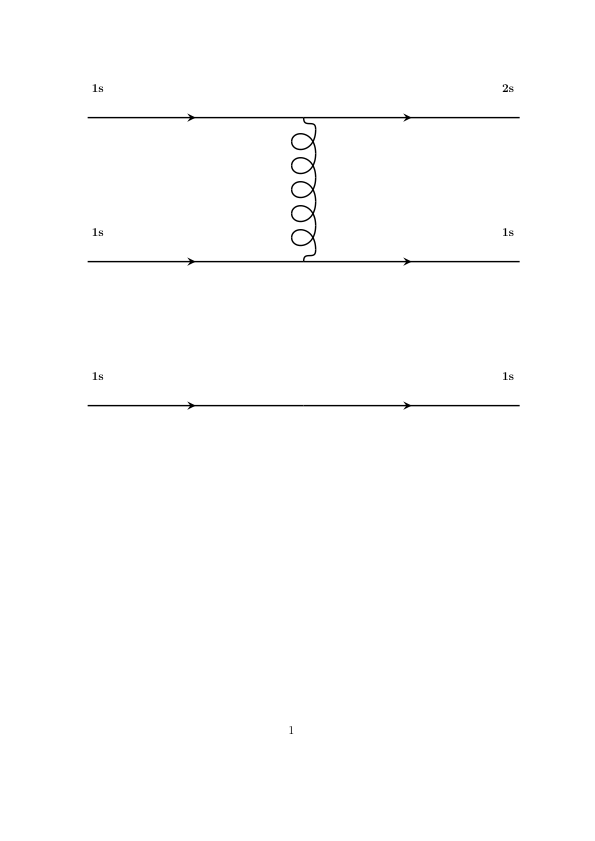}
  \hfill  
  \caption{Hyperfine one gluon exchange contribution leading to mixing between nucleon and $N^{*}$ states in the bag model.  }
\end{figure}

The reason for the small value of the mixing angle in the bag model is the small probability of `borrowing' enough energy from the vacuum to excite a quark into the $2s$ state. 
%In a similar fashion bag perturbation theory underestimates the size of the neutron charge radius by an order of magnitude \cite{CH80}. 
I note that this is in accord with Hazelton  \cite{Haz97}, who found that bag states with a $2s$ component contribute negligibly to the nucleon wavefunction in a configuration mixing 
calculation that included states up to 1.5 GeV heavier than the nucleon.
In contrast the model of \cite{IKK78} has the perturbation induced by the potential, and the overlap of oscillator wavefunctions is fairly large. 
Thus, while the hyperfine perturbation is greatly important in quark models, the mixing angle between nucleon and $N^{*}$ states is seen to differ in sign and magnitude between models.

\section{Nucleon - $N^{*}$ mixing in the Cloudy Bag Model}

From the previous section, we see that the hyperfine interaction does not necessarily generate large enough breaking of SU(6) symmetry to explain the behaviour of the 
valence distributions at large $x$.
The natural question then is whether there are other known interactions in quark models that also break SU(6) symmetry, and whether these interactions generate mixings, 
especially between the nucleon and the $N^{*}(70)$, that can give a good description of the valence distributions. 
An obvious candidate is the pion cloud of the nucleon. 
In the deep inelastic regime, the observed breaking of the Gottfried sum rule can be ascribed to the effects of the pion cloud \cite{NMC90, Th83, MTSS91}.
It has long been recognised that pion cloud effects improve many MIT bag model predictions, especially the (SU(6) breaking) neutron charge radius \cite{DeTar81, CBM81, Thomas84}.

Including pion loops in the model Hamiltonian generates a contribution to the mixing of the nucleon and the $N^{*}(70)$, as shown in figure 2. 
I choose to investigate this mixing in the cloudy bag model (CBM), as this model is renormalisable, and the model Hamiltonian can be generalized to include the resonance states 
$N^{*}(56)$ and $N^{*}(70)$.
The model Hamiltonian can be written as 
\bea
\hat{H}_{\text{CBM}}  =  \hat{H}_{\text{MIT}} + \hat{H}_{\pi} + \hat{H}_{\text{I}} % \equiv \hat{H}_{0}} +  \hat{H}_{I} 
\eea
where $\hat{H}_{\text{MIT}}$ is the MIT bag model Hamiltonian, extended as per the previous section to include the baryon states we are considering ($N, \Delta, N^{*}(56), N^{*}(70)$), 
$\hat{H}_{\pi}$ is the Hamiltonian for a free isovector pseudoscalar pion field $\vec{\phi}(\mathbf{k})$, 
and $\hat{H}_{\text{I}}$ now includes terms for all pertinent baryon - pion vertices.
%\bea
%\hat{H}_{\text{I}} = \sum_{\alpha, \beta, j} \int \frac{d \mathbf{k}}{(2\pi)^{3/2}} (v^{\alpha \beta}(\mathbf{k}) \alpha^{\dagger} \beta \phi_{j}(k) + \text{h.c.})
%\eea
%with operator $\alpha$ ($\alpha^{\dagger}$) annihilating (creating) a bare baryon.

\begin{figure}[bt]
\centering
  \includegraphics[width=8.6cm]{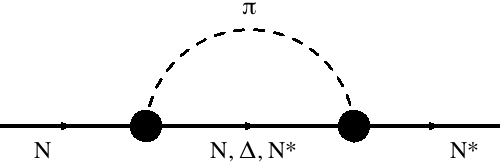}
  \hfill  
  \caption{One loop diagram generating mixing between nucleon and $N^{*}$ states in the CBM. }
\end{figure}

In the CBM we can define the (unrenormalised) quark-pion pseudoscalar surface coupling 
\bea
\frac{f_{q_{i} q_{f} \pi}}{m_{\pi}} (\vec{\sigma} \cdot {\mathbf{k}}) (\vec{\tau} \cdot \hat{\phi}) v(kR) 
= \frac{1}{2f_{\pi}} \int d {\mathbf{r}}  \, \bar{q}_{f}({\mathbf{r}}) \gamma_{5} \vec{\tau} q_{i}({\mathbf{r}}) \cdot \hat{\phi} \exp(i {\mathbf{k \cdot r}}) \delta(r - R).
\eea
As usual $f_{\pi} = 93 \: {\rm  MeV}$ is the pion decay constant, $\hat{\phi} = \vec{\phi} / |\phi|$ and $v(kR) = 3 j_{1}(kR)/(kR)$ is the CBM form factor.
Using the bag wavefunctions, equation (\ref{eq:bagwf}), then gives \cite{UDW83}
\bea
f_{1s \, 1s \, \pi} & = & \frac{m_{\pi}}{f_{\pi}}N_{1}^{2}R^{3} j_{0}^{2}(\Omega_{1}) = 0.486 \nonumber \\
f_{1s \, 2s \, \pi} & = & \frac{m_{\pi}}{f_{\pi}}N_{1}N_{2}R^{3} j_{0}(\Omega_{1})  j_{0}(\Omega_{2}) = -0.385 \nonumber \\
f_{2s \, 2s \, \pi} & = & \frac{m_{\pi}}{f_{\pi}}N_{2}^{2}R^{3} j_{0}^{2}(\Omega_{2}) = 0.302.
\eea

The coupling constants for the processes $B_{i} \rightarrow B_{f} \pi$ can now be found by taking matrix elements of the spin and isospin operators between appropriate eigenstates. 
For example the $N \rightarrow N^{*} \pi$ amplitude gives
\bea
\langle S_{z}=\frac{1}{2}, I_{3}=\frac{1}{2} | f^{0}_{NN^{*}\pi} \sigma_{z} \tau_{3} | S_{z}=\frac{1}{2}, I_{3}=\frac{1}{2} \rangle = 3 \langle N^{*} | f_{1s \, 2s \, \pi} \sigma_{z} \tau_{3} b^{\dagger}_{2s} b_{1s} | N \rangle
\eea
where the operators on the right-hand side are understood to be operating on the third quark. 
The various (unrenormalised) coupling constants are given in table 1. 
I note that the expression I obtain for $f^{0}_{N^{*}(70) N^{*}(70) \pi}$ differs from that in \cite{UDW83}.

\begin{table}
\begin{tabular}{|c|c|c|} \hline
$f^{0}_{N N \pi}$   & $\frac{5}{3} f_{1s \, 1s \, \pi} $ & 0.81 \\ 
$f^{0}_{N \Delta \pi}$   & $\sqrt{\frac{72}{25}} f^{0}_{N N \pi} $   & 1.37 \\
$f^{0}_{N N^{*}(56) \pi}$   & $\frac{5 \sqrt{3}}{9} f_{1s \, 2s \, \pi} $   & -0.369 \\ 
$f^{0}_{N N^{*}(70) \pi}$   & $-\frac{4 \sqrt{3}}{9} f_{1s \, 2s \, \pi} $   & 0.295 \\  \hline
$f^{0}_{\Delta \Delta \pi}$   & $\frac{4}{5} f^{0}_{N N \pi} $   & 0.65 \\
$f^{0}_{\Delta N^{*}(56) \pi}$   & $\frac{4 \sqrt{6}}{3} f_{1s \, 2s \, \pi} $   & -1.25 \\ 
$f^{0}_{\Delta N^{*}(70) \pi}$   & $\frac{4 \sqrt{6}}{3} f_{1s \, 2s \, \pi} $   & -1.25 \\  \hline 
$f^{0}_{N^{*}(56) N^{*}(56) \pi}$ & $\frac{5}{9}( 2f_{1s \, 1s \, \pi} + f_{2s \, 2s \, \pi} )$ & 0.71 \\ 
$f^{0}_{N^{*}(56) N^{*}(70) \pi}$ & $\frac{4}{9}( f_{1s \, 1s \, \pi} - f_{2s \, 2s \, \pi} )$ & 0.082 \\ 
$f^{0}_{N^{*}(70) N^{*}(70) \pi}$ & $\frac{1}{9}( 5f_{1s \, 1s \, \pi} - 4f_{2s \, 2s \, \pi} )$ & -0.048 \\  \hline
\end{tabular}
\caption{Baryon-Baryon-Pion coupling constants in the cloudy bag model in terms of the quark-quark-pion coupling constants}
\end{table}

The one loop contributions to the self-energy of the nucleon and $N^{*}$ states can now be calculated. 
In the static baryon limit we have 
\bea
\delta M_{fi} = \sum_{B} T_{B} \,{\rm P} \int_{m_{\pi}}^{\infty} d \omega \, f^{0}_{B_{f} B \pi} f^{0}_{B_{i} B \pi} \frac{k^{3} v^{2}(kR)}{(2\pi)^{2} m_{\pi}^{2} (E - M_{B} - \omega)} \left|_{E= M_{B_{i}} }\right.
\label{eq:loops}
\eea
where $B = N, \Delta, N^{*}(56)$ and $N^{*}(70)$, and I use the physical masses of the nucleon and delta and also the $N(1440)$ and $N(1710)$ masses for the $N^{*}$ states.
%the zeroth order bag model mass (equation \ref{eq:mass}) of the $N^{*}$ states. 
The isospin loop factor $T_{B}$ is 3 for $N$ and $N^{*}$ states, and 1/3 for $\Delta$ baryons in the loop.
The CBM form factor cuts off the integral at $\omega_{{\rm max}} = (k_{1}^{2} + m_{\pi}^{2})^{1/2}$ where $k_{1}R \approx 4.49 $ is the first zero of $v(kR)$.

For a bag radius of 5 GeV$^{-1}$ (0.99 fm) I find that the splitting between the nucleon and the $\Delta$ from the one pion loops is 53 MeV, which is similar to the splitting in the analysis of 
Myhrer and Thomas  \cite{MT08}, and is consistent with analysis of lattice simulations \cite{YLTW02}. 
A small value of the one pion loop contribution to the $N - \Delta$ splitting means that we can consider both pion loop and one gluon exchange contributions independently with only a 
small risk of double counting. 
So a consistent picture emerges here, where the OGE is responsible for around 80\% of the $N - \Delta$ splitting and one pion loops are responsible for the remaining 20\%, 
which decreases the value of the strong coupling constant in the model. 
As both these effects scale as $R^{-1}$, this picture should remain valid for a reasonable range of model parameters, 
however, for small bag radius the one pion loop contribution to the $N - \Delta$ splitting may become significantly larger than 100 MeV, and the absolute shift in both masses becomes large. 
It is possible that renormalisation of the $NN\pi$ coupling constant may ameliorate this. 

Summing the contributions in equation (\ref{eq:loops}) yields the one pion loop shifts in the nucleon - $N^{*}$ mass matrix
\bea
\delta \bar{M}^{\pi} = 
	%\begin{array}{c} N \\ N^{*}(56) \\ N^{*}(70) \end{array}
	\left( 
	\begin{array}{ccc}
	-126 & 77 & 24 \\
	77 & -69 & -12 \\
	24 & -12 & -12
	\end{array}
	\right)
	\left[ \frac{5 \; {\rm GeV}^{-1}}{R} \right] {\rm MeV}
\label{eq:pishift}
\eea
where the rows of the matrix correspond to the shifts in the nucleon mass,  $N^{*}(56)$ mass and  $N^{*}(70)$ mass respectively. 
These shifts are similar in magnitude to shifts in the $N^{*}$ masses from the OGE \cite{CH80}, so both contributions should be considered.
The OGE shifts in the mass matrix can be written 
\bea
\delta \bar{M}^{{\rm OGE}} = 
	%\begin{array}{c} N \\ N^{*}(56) \\ N^{*}(70) \end{array}
	\left( 
	\begin{array}{ccc}
	-0.531 & -0.004 & -0.004 \\
	-0.004 & -0.422 & 0.109 \\
	-0.004 & 0.109 & -0.223
	\end{array}
	\right) \kappa
\label{eq:ogeshift}
\eea
where the strength of the OGE is set by requiring that the nucleon - delta mass splitting is reproduced correctly taking into account the pion loop contribution: 
\bea
\kappa = 0.276 - 0.049 \left[ \frac{5 \; {\rm GeV}^{-1}}{R} \right] {\rm GeV}.
\eea
I note that the off-diagonal pionic and gluonic contributions are opposite in sign, which reduces the mixing. 

For a bag radius of 5 GeV$^{-1}$ I obtain the complete mass matrix
\bea
\bar{M} & = & \bar{M}_{N}^{0} I + \delta \bar{M}^{\pi} + \delta \bar{M}^{{\rm OGE}} \nonumber \\
& = & 	\left( 
	\begin{array}{ccc}
	0.832 & 0.076 & 0.023 \\
	0.076 & 1.367 & 0.012 \\
	0.023 & 0.012 & 1.469
	\end{array}
	\right) {\rm GeV}.
\label{eq:masses}
\eea 
The physical masses are found by diagonalizing the mass matrix, resulting in 
\bea
M_{N} & = & 0.821 \:  {\rm GeV} \nonumber \\
M_{N^{*}(56)} & = & 1.375 \:  {\rm GeV} \nonumber \\
M_{N^{*}(70)} & = & 1.472 \:  {\rm GeV}
\eea
which are similar to the masses found by Umland, Duck and von Witsch \cite{UDW83}.
The mixing angles between the states are obtained from the respective eigenvectors of the mass matrix. 
For the nucleon we have
\bea
| \tilde{N} \rangle = 0.990 |N \rangle - 0.136 | N^{*}(56) \rangle - 0.033 | N^{*}(70) \rangle.
\eea
This mixing angle $\theta = -0.033$ is larger in magnitude than the very small mixing coming from the OGE in the model, however, it is considerably smaller than the mixing angle 
in the Isgur and Karl model. 
If the OGE contribution is ignored, the mixing angle becomes $\theta = -0.044$.

I have made no attempt here to fit the masses of the nucleon and the $N^{*}$ system to the known physical masses, however, the masses determined in equation (\ref{eq:masses}) are 
all smaller than the known values. 
Decreasing the bag radius $R$ usually increases the mass of bag model states, though it can be seen from the top left entries of the $\delta \bar{M}^{\pi}$ and $\delta \bar{M}^{{\rm OGE}}$ matrices, 
that such a decrease will also result in a larger (negative) shift in the nucleon mass. 
However, decreasing the bag radius increases the splitting between the $N^{*}(56)$ and $N^{*}(70)$ states, which improves agreement with the physical values.
Choosing a bag radius of 4 GeV$^{-1}$ (0.79 fm) results in mass eigenvalues of 
\bea
M_{N} & = & 0.822 \:  {\rm GeV} \nonumber \\
M_{N^{*}(56)} & = & 1.49 \:  {\rm GeV} \nonumber \\
M_{N^{*}(70)} & = & 1.60 \:  {\rm GeV}
\eea
and gives a mixing angle of $\theta = -0.036$.  
Taking into account shifts due to both pion loops and the OGE, the mixing angle shows little dependence on bag radius $R$.

\section{Perturbed Valence Quark Distributions}

In order to gain greater insight into the behaviour of the distortion term $w(x)$ arising from the mixing of the $N^{*}(70)$ wavefunction with the nucleon wavefunction, 
I attempt to calculate it using the MIT bag model and the approach of  \cite{SST91}. 
This will enable a direct comparison between the perturbative approach and the approach using an exact hyperfine basis in the same hadron model.
I note a previous calculation using the perturbative approach with the non-relativistic quark model \cite{DMZ94}. 
In this model the relativistic corrections come from the Melosh transformation from the rest frame to the infinite momentum frame.

To calculate the valence quark distributions in the bag model I start from the expression \cite{SST91, Jaffe83, ST89}
\bea
q(x) = p^{+} \sum_{n} \delta(p^{+}(1-x) - p_{n}^{+}) \langle p | \Psi_{+}^{\dagger} (0) |n \rangle \langle n | \Psi_{+} (0) |p \rangle
\eea
where the initial eigenstate $|p \rangle$ now has the hyperfine perturbed spin-flavour wavefunction given by equation (\ref{eq:hpwf}).
%To form the momentum eigenstates I take a Peierls-Yoccoz projection \cite{PY57} of the MIT bag wavefunctions for the proton, 
%$N^{*}$ and the intermediate diquark state $|n \rangle$. 
To form the momentum eigenstates I use the non-relativistic Peierls-Yoccoz projection \cite{PY57} of the MIT bag wavefunctions for the proton, 
$N^{*}$ and the intermediate diquark state $|n \rangle$ - see \cite{SST91, AS93} for the limitations of using this projection.
This gives the leading order correction to the valence quark distribution as 
\bea
w(x) & = & -  \frac{\sqrt{2} M}{(2\pi)^{3}} \int d \mathbf{p}_{n}\: \delta(M(1-x) - p_{n}^{+})  \frac{|\phi_{2}(\mathbf{p}_{n})|^{2}}{\phi_{3}(0) \tilde{\phi}_{3}(0)}  \nonumber \\
& &\;\; \times \left( \tilde{\Psi}_{1s+}^{\dagger}(\mathbf{p}_{n})  \tilde{\Psi}_{2s+}(\mathbf{p}_{n}) + \tilde{\Psi}_{2s+}^{\dagger}(\mathbf{p}_{n})  \tilde{\Psi}_{1s+}(\mathbf{p}_{n}) \right)
\label{eq:wdist}
\eea
where $\tilde{\Psi}_{1s/2s}(\mathbf{p})$ is the Fourier transform of the $1s$ or $2s$ bag wavefunction,
the projector over `good' components of the wavefunction on the light-cone is 
\bea
P_{+} = \frac{1}{2} \gamma^{-} \gamma^{+} = \frac{1}{2} (I + \alpha^{3}), 
\eea
and $\phi_{n}(\mathbf{p})$ is the Fourier transform of the $n$ quark Hill-Wheeler overlap for all quarks in the ground state, 
while $\tilde{\phi}_{3}(\mathbf{p})$ is the same but with 1 quark in the $2s$ state.

The final parameters required for the calculation are the proton mass and bag radius, and the mass of the intermediate diquark state $M_{n} = \sqrt{p_{n}^{2}}$. 
I assume that the mixing of the nucleon and $N^{*}$ states reproduces the physical proton mass 0.940 GeV, 
and take the bag radius as that used earlier, $R =  5.0 \; {\rm GeV}^{-1}$ (0.99 fm). 
As the mixing angle $\theta $ is small, the final bag radius can be taken as equal to this value.  
The intermediate state mass is chosen to be 75\% of the nucleon mass, similar to the calculation of \cite{SST91}, though I note that other 
choices could be made here. 

In figure 3a I show the calculated distribution $w(x)$ for bag radius $R = 5.0 \; {\rm GeV}^{-1}$, and intermediate state masses $M_{n} / M = 0.675, 0.75 $ and $0.825$, 
{\em i.e.} varying by $\pm10\%$. 
Increasing (decreasing) $M_{n}$ appears to shift the distribution up (down).
The integral of $w(x)$ varies from $0.1 - 0.2$.
While this is not exactly zero, as expected, when multiplied by the small value of the mixing angle it is much smaller than the first moment of the calculated 
valence distribution $ \langle v(x) \rangle = 0.87$.
I remind the reader that that this value is not unity in the bag model calculation because contributions at negative $x$ have been ignored \cite{ST89, SST91}.
From the figure we can expect that the contributions to $w(x)$ from the negative $x$ region will be negative and at least partially cancel the small positive 
value calculated over positive $x$.
I also note that the Peierls-Yoccoz projection is non-relativistic, and is unlikely to be valid for $x$ much larger than 0.7, as above this value the integrand  
in equation (\ref{eq:wdist}) is dominated by relativistic momenta.
This means that I cannot determine whether the very small negative contribution to $w(x)$ above $x \sim 0.75$ seen in the figure is a true feature of the distribution or an 
artefact of the approximations used here. 
I note that in calculations of $w(x)$ at different bag radii and different $M_{n}/M$ this feature is not always present. 
Unfortunately, this makes a calculation of the ratios $d(x) / u(x)$ and $F_{2}^{n}(x) / F_{2}^{p}(x)$ at large $x$ meaningless, as these ratios change dramatically when the model 
parameters are varied. 

We see that $w(x) $ peaks around $x = 0.5 \sim 0.6$, which is a consequence of the $2s$ quark in the $N^{*}(70)$ bag wavefunction carrying a large proportion of the energy-momentum. 
This can be compared with calculations of proton valence distributions in the bag model, which peak in the region $x = 0.3 \sim 0.4$ \cite{SST91}. 
If the $N^{*}(70)$ had a large pentaquark component, we would expect the contributions to the valence quark distributions to peak around $x = 0.2$, and any effect at large $x$ to be small.

\begin{figure}[bt]
\centering
  \includegraphics[width=8.6cm]{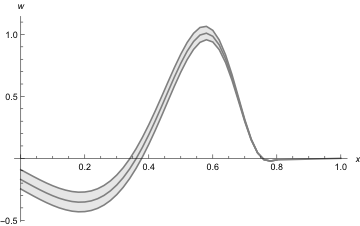}
  \includegraphics[width=8.6cm]{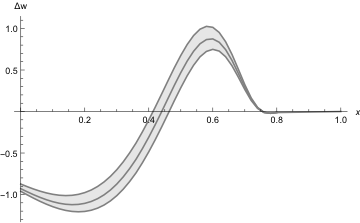}
  \hfill  
  \caption{The first order contributions $w(x)$ and $\Delta w(x)$ to the unpolarised and polarised valence quark distributions for bag radius $R = 5.0 \; {\rm GeV}^{-1}$.  
  The central lines show the distributions calculated for intermediate state mass 75\% of the proton mass, while the hatching shows how the distributions changes as
  the intermediate state mass is varied by $\pm 10 \%$.  }
\end{figure}

In figure 3b I show the calculated spin dependent distribution $\Delta w(x)$ for the same parameters as the spin independent distribution. 
In a relativistic model, such as the bag, there is no relativistic correction term $c_{A}(x)$, 
however, relativity does affect the spin dependent distributions slightly differently to the spin independent distributions \cite{SST91} due the properties of the helicity projection 
operators $ P^{\ua \da} = (1 \pm \gamma_{5})/2$.
Thus the distributions required are written in terms of the spin dependent valence distribution $\Delta v(x)$ and a spin dependent leading order correction $\Delta w(x)$. 
It is expected that the first moment of $\Delta v(x)$ will be less than that of the spin independent $v(x)$, while the first moment of $\Delta w(x)$ should be small, 
but not necessarily zero.
I obtain
\bea
\Delta u(x) & = & \frac{4}{3}\Delta v(x) - \frac{5 \theta}{3}  \Delta w(x)  \\ 
\Delta d(x) & = & -\frac{1}{3}\Delta v(x) - \frac{\theta}{3} \Delta w(x).
\eea
In order to calculate the proton and neutron asymmetries I have only used $x \leq 0.7$ as the sign of $\Delta w(x)$ in this region is stable when the calculation parameters are varied.

In figure 4 I show the spin dependent valence distributions calculated to first order with bag radius $5.0 \; {\rm GeV}^{-1}$ and intermediate state mass 75\% of the proton mass, 
for two values of the mixing angle $\theta$ between the SU(6) nucleon and $N^{*}(70)$ states. 
I have used $\theta = -0.04$ corresponding to the value obtained earlier using the CBM, and $\theta = -0.23$ which is the value used by Isgur \cite{Isgur99}. 
These distributions are also compared with those calculated using the bag model in an exact hyperfine basis \cite{SST91} for the same bag radius and hyperfine splitting of 200 MeV. 
While the $\Delta d(x)$ is similar for all three calculations, we see that the $\Delta u(x)$ distribution is much larger in the perturbation approach. 
We can understand the reason for these similarities and differences between these distributions in terms of the intermediate state mass $M_{n}$ in each case.
For the $\Delta d(x)$ distribution calculated in the exact hyperfine basis, the distribution comes entirely from terms where the intermediate state is in a triplet state, which has a mass 
only 50 MeV larger than the intermediate state mass used for the perturbative calculation. 
The relatively small difference in intermediate state mass between that calculation and the zeroth order term in the perturbative calculation then leads to only small changes in the distribution.
On the other hand, in the exact hyperfine basis $\Delta u(x)$ is dominated by terms where the intermediate state is a singlet, with mass 150 MeV less than the mass being used in 
the perturbative calculation.
This much larger difference in intermediate state masses gives rise to most of the differences seen in figure 4.

\begin{figure}[bt]
\centering
  \includegraphics[width=8.6cm]{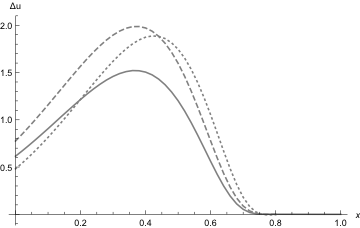}
  \includegraphics[width=8.6cm]{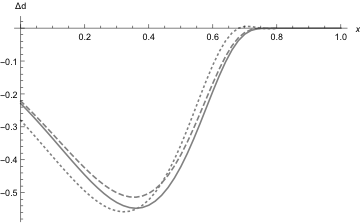}
  \hfill  
  \caption{Comparison of spin dependent valence up and down distributions calculated in the bag model with $R = 5.0 \; {\rm GeV}^{-1}$. 
  The solid line is the calculation in an exact hyperfine basis while the dashed and dotted lines are the 1st order distributions with mixing angle $\theta = -0.04$ and $-0.23$ respectively.}
\end{figure}

Note that the valence distributions are calculated without including the effects of the external current coupling to pion loops \cite{Sull72}. 
Such terms are included in the meson cloud model (MCM) \cite{HSS96, ST97}, and are known to give important corrections to the Gottfried and Bjorken sum rules. 
However, these corrections are for the most part at low $x$ as the pion momentum distribution $f_{\pi}(y)$ is small at large $y$, 
so I have not considered them here.

In figures 5 and 6 I show the proton and neutron asymmetries calculated to first order with bag radius $R = 5.0 \; {\rm GeV}^{-1}$ and intermediate state mass 75\% of the proton mass. 
Again, I show these for two values of the mixing angle $\theta$.
The proton asymmetry is compared with figure 1(a) of \cite{Isgur99}, showing experimental data in the range $ 0.3 \leq x \leq 0.8$ and the band of rough parametrizations 
calculated in that paper. 
For the small magnitude of the mixing angle, the agreement with the data is fairly good, whereas for the larger magnitude of the mixing angle the agreement is good at small $x$, 
but appears to rise faster than the data at large $x$.

The neutron asymmetry is compared with a simplified version of figure 1 of \cite{E9904} showing large $x$ data from the E99-117 experiment at Jefferson Lab, as well as earlier 
experimental data, the band of parametrizations calculated in \cite{Isgur99} and some other model calculations and parametrizations. 
In both cases the agreement with data is reasonably good in the range $ 0.3 \leq x \leq 0.7$, however the calculation with $\theta = -0.04$ is rather flat, and not in particularly good 
agreement at low values of $x$.
While this calculation is performed at the model scale, somewhere in the vicinity of $\mu^2 = 0.2 - 0.3 \: {\rm  GeV}^{2}$ and is not evolved to experimental scales, evolution effects 
are small at at both leading and next-to-leading order QCD for the asymmetries, so I expect the level of agreement seen here to hold at experimental scales. 

It is interesting that both values of the mixing angle between the nucleon and the $N^{*}(70)$ used here can give reasonable descriptions of the proton and neutron asymmetries. 
This suggests that the mixing is a viable hypothesis for the observed SU(6) symmetry breaking in the proton and neutron asymmetries and  spin dependent valence quark distributions. 
The non-relativistic quark  model suggests that this mixing is due to the hyperfine interaction, whereas the cloudy bag model suggests that both hyperfine and pion loop effects are 
responsible. 
These calculations give hope that both of these models can be tested in a systematic manner.

\begin{figure}[bt]
\centering
  \includegraphics[width=8.6cm]{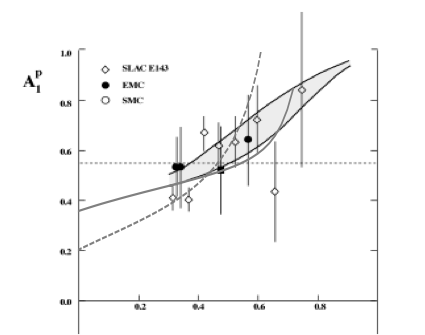}
  \hfill  
  \caption{The first order corrected proton asymmetry. The solid and dashed lines are the calculated asymmetries for $\theta = -0.04$ and $-0.23$ respectively, 
  and the shaded band is the calculation of \cite{Isgur99}.}
\end{figure}

\begin{figure}[bt]
\centering
  \includegraphics[width=8.6cm]{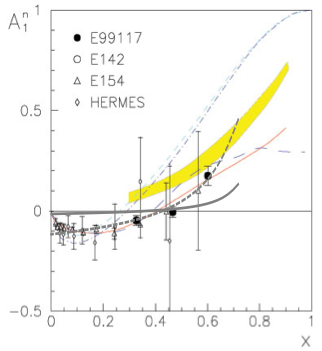}
  \hfill  
  \caption{The first order corrected neutron asymmetry. The thick solid and dashed lines are the calculated asymmetries for $\theta = -0.04$ and $-0.23$ respectively, 
  and the shaded band is the calculation of \cite{Isgur99}. 
  Other curves are: thin solid line - NLO  parametrisation \cite{LSS02}; long dashed line - statistical quark model \cite{BSB02}; short dashed and dash-dot lines - pQCD parametrisations of  $g_{1}^{n}/F_{1}^{n}$\cite{LSS98, BBS95}. }
\end{figure}

\section{Perturbative QCD}

In perturbative QCD, scattering from a quark polarised in the opposite direction to the proton polarisation is suppressed compared to scattering when the quark and proton polarisations are aligned. 
This arises because the former requires the exchange of a longitudinal gluon when the spectator di-quark $(qq)_{S}$ has both spins aligned \cite{FJ75}. 
Components of the proton wavefunction with di-quark spin component $S_{z} = 1$ are suppressed by a factor $(1 - x)$ relative to those with $S_{z} = 0$. 
Brodsky, Burkhardt and Schmidt \cite{BBS95} show that the large $x$ behaviour of quark distributions is given by the rule:
\bea
q^{\ua \da} (x) \rightarrow (1- x)^{2n-1+ \Delta S_{z}} & x \rightarrow 1
\eea
where $n$ is the minimum number of spectator quarks (2 for valence distributions) and $ \Delta S_{z}$ is the difference between the proton helicity and the helicity of the struck quark.
Similar to the hyperfine basis, the proton wavefunction can be written
\bea
|p \ua \rangle & = & \frac{1}{\sqrt{2}} |u \ua \rangle (ud)_{S=0} + \frac{1}{\sqrt{18}} |u \ua \rangle (ud)_{S=1,S{z}=0} - \frac{1}{3} |u \da \rangle (ud)_{S=1,S{z}=1} \nonumber \\
&&\ - \frac{1}{3} |d \ua \rangle (uu)_{S=1,S{z}=0} - \frac{\sqrt{2}}{3} |d \da \rangle (ud)_{S=1,S{z}=1}
\label{eq:pqcdwf}
\eea
where I now suppress the spatial part of the wavefunction.
From this we obtain as $x \rightarrow 1$
\bea
\frac{\Delta d}{d} \rightarrow 1,   \frac{\Delta u}{u} \rightarrow 1,  
\eea
and for the structure functions and asymmetries
\bea
\frac{F_{2}^{n}(x)}{F_{2}^{p}(x)} & \rightarrow & \frac{3}{7} \\
A_{1}^{p} & \rightarrow &  1  \\ 
A_{1}^{n} & \rightarrow & 1.
\eea

In model calculations the correct power of $(1 - x)$ can be input by hand, however I note that care must be taken with the wavefunction in equation (\ref{eq:pqcdwf}) 
to avoid it being inconsistent with the Pauli principle, similar to the earlier discussion following equation (\ref{eq:2chf}) of the solutions in a hyperfine basis. 
The correction due to pQCD can be considered to introduce a term in the Hamiltonian which has the effect of breaking the SU(2) symmetry of the hyperfine basis. 
Similar to above, the pQCD Hamiltonian could be treated as a perturbation to a model Hamiltonian describing hadron structure. 
This approach would require a careful analysis to ensure that the perturbation remains small over the majority of the kinematic domain.

Another important question is the need to define where the large $x$ regime lies for pQCD. 
A rise in the neutron asymmetry would be a definite sign of this physics being important.
There is a strong hint of this in the data point at $x = 0.6$ from the E99-117 experiment at Jefferson Lab ($A_{1}^{n}(x = 0.6) = 0.175 \pm 0.048$ (stat) $+ 0.026 -0.028$ (sys)) \cite{E9904},
but further data to confirm this is needed. 
There has been progress in determining the ratio of unpolarized quark distributions at large $x$. 
Experimental data from the BoNuS experiment has measured $F_{2}^{n} / F_{2}^{d}$ with high precision up to $x = 0.8$ \cite{Bonus12} 
and shows that this ratio falls well below the SU(6) value at large $x$, and is probably larger than 0.25 as $x \rightarrow 1$. 
Also the D$\varnothing$ experiment measuring charged lepton and W asymmetries) \cite{Dzero13} is able to strongly constrain the ratio $d/u$ at large $x$.
The CTEQ - Jefferson lab collaboration obtain \cite{CJ15}
\bea
\frac{d}{u} \rightarrow 0.09 \pm 0.03
\eea
as $x \rightarrow 1$, compared with the pQCD value of $1/5$. 
As only pQCD predicts a non-zero value of this ratio, this would appear to be a good indication that $S_{z} = 1$ components of the proton wavefunction are indeed suppressed.

\section{Discussion and Conclusions}

In this paper I have re-examined the problem of SU(6) symmetry breaking of the valence quark distributions of the proton, particularly the spin dependent distributions, 
in the context of quark models of the nucleon. 
The effects of the hyperfine interaction in quark models, which is often thought to be the main source of the symmetry breaking, were studied in some detail. 
I showed that calculations of valence quark distributions using a basis of hyperfine eigenfunctions which break SU(2) in the quark model do not require any special treatment 
\emph{viz} the Pauli exclusion principle, contrary to what has been previously claimed.
In addition, I have examined the perturbative approach to the hyperfine splitting, and shown how the distortions (or 1st order correction terms) $w(x)$ and $\Delta w(x)$ 
can be calculated in terms of the interference matrix elements between the ground state wavefunction and the excited states, particularly the mixed symmetry $N(70, +)$ state.
In the bag model I have shown that the hyperfine contribution to the interference matrix elements is small, and that pion loop contributions should also be taken into account to 
determine the mixing angle. 
With both effects included, I found a mixing angle of $\theta \approx -0.04$, which is smaller than the value of -0.23 found in the hyperfine-perturbed non-relativistic quark model. 
However, distributions calculated with both these values of the mixing angle were able to give reasonable agreement with experimental data on the proton and neutron asymmetries 
over a large range up to $x \approx 0.7$.
This leads to the interesting possibility that it is not the hyperfine interaction \emph{per se} that leads to SU(6) symmetry breaking, but more generally an interaction that allows the 
SU(6) nucleon to mix with the $N^{*}(70)$ state can lead to the observed SU(6) breaking in the valence quark distributions.

I have not discussed models where the spin dependent forces between quarks are due to the effects of instantons, or where the axial anomaly plays a large role in suppressing 
the total spin $\Sigma$ carried by quarks.
Such models generally require $\Delta G$ to be large for the proton. 
RHIC experiments appear to show that  $\Delta G$ is small \cite{RHIC07}, so the effects predicted in these models will be small in the proton.

The experimental situation has improved in this century.
As discussed above, BONuS and Fermilab collider experiments have improved the knowledge of unpolarized distributions at large $x$. 
While $F_{2}^{n} / F_{2}^{p}$ is clearly below the SU(6) value, it is not clear whether the ratio will equal $\frac{1}{4}$ or $\frac{3}{7}$, or fall somewhere between, as $x \rightarrow 1$.
There has been a great deal of effort in polarised experiments at Compass, Hermes and Jefferson Lab.
However, most of this has been at $x < 0.5$, and has little implication for large $x$ physics.
The exception has been the E99-117 experiment, which shows that SU(6) is broken for the neutron asymmetry, 
and that $A_{1}^{n}$ is positive for $x > 0.5$ in line with the expectations of SU(6) symmetry breaking.

There are a number of interesting possibilities for follow-up work to this paper. 
It would be interesting to see whether lattice simulations are able to find evidence of the $N(70, +)$ mixing with the nucleon, and what range of mixing angles these simulations indicate. 
Clarification of the physics of the positive parity nucleon resonances, particularly the $N(1440) $ and $N(1710)$, would also be useful. 
The quark content of these resonances, and their correct SU(6) wavefunctions, is still a matter of debate, which could have strong consequences for understanding the breaking of SU(6) in the proton. 

The calculations presented here should be repeated including the symmetric $N(56, +)$ component of the nucleon wavefunction to check that the symmetry breaking is a general feature of 
the perturbed wavefunction. 
A calculation using the non-relativistic quark model \cite{DMZ94} found that including the symmetric state in the wavefunction of the nucleon tended to result in smaller asymmetries. 
In addition, more detailed calculations of quark distributions, covering smaller $x$ than considered here, can investigate the interplay between the SU(6) breaking manifest in the meson cloud model \cite{ST97} 
at low $x$ and the symmetry breaking from $N^{*}(70)$ mixing. 
Evolving these quark distributions to experimental scales will enable a thorough comparison with data and better constraints on model parameters and, hopefully,  the mixing angle $\theta$.
I note that the bag model calculations in this paper, and similar calculations of valence quark distributions in non-relativistic models have problems at large $x$, where one quark must be 
highly relativistic. 
For quark models to offer greater insight into large $x$ physics it is vital that fully relativistic methods of calculating quark distributions be developed. 
Unfortunately, progress in this area is slow.
Calculations of generalised parton distributions (GPDs) can give further insight into the breaking of SU(6). 
In particular, moments of various GPDs are related to quark angular momentum and orbital angular momentum $L$, though there is some debate over the interpretation of these.
Both relativistic corrections and hyperfine interactions predict that $L$ is positive, which can be verified by experiment.

%%%%%%%%%%%%%%%%%%%%%%%%%%%%%%%%%%%%%%%%%%%%%%%%
%% BACKMATTER
%%%%%%%%%%%%%%%%%%%%%%%%%%%%%%%%%%%%%%%%%%%%%%%%

\section*{Acknowledgments}
I am grateful to Frank Close for help with some early sections of this work, 
and to Fu-guang Cao and Tony Thomas, who read early versions of this manuscript and generously offered a number of helpful insights.



\bibliographystyle{aipproc}   % if natbib is available

%\bibliography{sample}

%  \typeout{******************************************}
 % \typeout{** Please run "bibtex \jobname" to optain}
%  \typeout{** the bibliography and then re-run LaTeX}
%  \typeout{** twice to fix the references!}
%  \typeout{******************************************}

\end{document}